\documentclass[letter]{aa} 

\usepackage{natbib}                     
\usepackage{amsmath}                    
\usepackage{graphicx}                   
\usepackage[varg]{txfonts}              
\usepackage[english]{babel}             
\usepackage{nameref}                    
\usepackage[normalem]{ulem}             
\usepackage{paralist}                   
\usepackage{multirow}

\begin{document}

  \title{Constraining planet structure from stellar chemistry: the cases of CoRoT-7, Kepler-10, and Kepler-93\thanks{Based on archival data obtained with the SOPHIE (1.93-m telescope OHP observatory), HARPS (3.6-m ESO, La Silla-Paranal Observatory), and HARPS-N (TNG telescope, La Palma) spectrographs.}}

  \author{N. C. Santos\inst{1,2}
          \and V. Adibekyan\inst{1}
          \and C. Mordasini\inst{3}
          \and W. Benz\inst{3}
          \and E. Delgado-Mena\inst{1}
          \and C. Dorn\inst{3}
          \and L. Buchhave\inst{4,5}
          \and P. Figueira\inst{1}
          \and A. Mortier\inst{6}
          \and F. Pepe\inst{7}
          \and A. Santerne\inst{1}
          \and S. G. Sousa\inst{1}
          \and S. Udry\inst{7}
         }

  \institute{
          Instituto de Astrof\'isica e Ci\^encias do Espa\c{c}o, Universidade do Porto, CAUP, Rua das Estrelas, 4150-762 Porto, Portugal
          \and
          Departamento de F\'isica e Astronomia, Faculdade de Ci\^encias, Universidade do Porto, Rua Campo Alegre, 4169-007 Porto, Portugal
         \and
          Physikalisches Institut, University of Bern, Sidlerstrasse 5, 3012, Bern, Switzerland
        \and
        Harvard-Smithsonian Center for Astrophysics, Cambridge, Massachusetts 02138, USA
        \and
        Centre for Star and Planet Formation, Natural History Museum of Denmark, Univ. of Copenhagen, 1350 Copenhagen, Denmark
        \and
        SUPA, School of Physics and Astronomy, University of St Andrews, St Andrews KY16 9SS, UK
        \and
        Observatoire de Gen\`eve, Universit\'e de Gen\`eve, 51 ch. des Maillettes, CH-1290 Sauverny, Switzerland
}

  \date{Received date / Accepted date }
  \abstract
  {}
  {We explore the possibility that the stellar relative abundances of different species can be used to constrain the bulk abundances of known transiting rocky planets.}
  {We use high resolution spectra to derive stellar parameters and chemical abundances for Fe, Si, Mg, O, and C in three stars hosting
  low mass, rocky planets: CoRoT-7, Kepler-10, and Kepler-93. {These planets follow the same line along the mass-radius diagram, pointing toward a similar composition.}
  The derived abundance ratios
  are compared with the solar values. With a simple stoichiometric mode, we
  estimate the iron mass fraction in each planet, assuming stellar composition.}
  {We show that in all cases, the iron mass fraction inferred from the mass-radius relationship seems to be in good
  agreement with the iron abundance derived from the host star's photospheric composition.}
  {The results suggest that stellar abundances can be used to add constraints on the composition
  of orbiting rocky planets.}


  \maketitle
   
  \section{Introduction}                                        \label{sec:Introduction}

The study of chemical abundances in stars with planetary companions proved to
be fundamental for our understanding of the formation of planetary systems as a whole \citep[][]{Ida-2004b,Mordasini-2012}. 
Important clues for the processes of planet formation and evolution have been brought by the discovery that the frequency of giant planets is a strong function
of stellar metallicity \citep[e.g.][]{Santos-2004b,Fischer-2005}, as well
as by the fact that such a correlation does not seem to be present for stars hosting lower mass 
planets \citep[][]{Sousa-2011,Buchhave-2012} -- see, however, discussions in 
\citet[][]{Adibekyan-2012}  and \citet[][]{Wang-2015}. Also,
the stellar chemical composition seems to be related to the
structure of the planets that were formed: the heavy 
element content of a giant planet seems to correlate with the host star metallicity \citep[][]{Guillot-2006,Fortney-2007}.
The whole planetary architecture (e.g., orbital periods and eccentricities) may be related to the presence of
heavy elements in the stellar host \citep[][]{Beauge-2013,Dawson-2013,Adibekyan-2013}.

Specific relative abundances of chemical species may also be reflected in the bulk composition of the planets \citep[e.g.,][]{Grasset-2009}.
Different chemical abundances in the disk
may result in the formation of planets that have different composition and
structure \citep[][]{Carter-Bond-2012}, something that may even change their 
habitability potential \citep[][]{Noack-2014}.
Understanding whether the relative chemical abundances we measure in the photosphere of the host star are
related to the relative bulk composition of its orbiting planet(s) may thus
provide valuable clues for modeling planet structure once precise masses and radii are measured.

Recently, \citet[][]{Dressing-2015} has shown that five known rocky
planets (\object{Kepler-10b}, \object{Kepler-36b}, \object{Kepler-78b}, \object{Kepler-93b}, and \object{CoRoT-7b}) 
with precise measurements of the mass (relative uncertainties less than 20\%) and radius all seem to fall along the same line in the mass-radius diagram: the area corresponding to an approximately Earth-like composition with a silicate mantle and an iron core of a similar mass fraction. Their best fit-model is comprised of 17\% iron and 83\% silicate mantle (MgSiO3), using a simple two-component interior structure model \citep{Zeng-2013}. {This is within 20\% of the iron mass fractions estimates of Earth, Venus, and Mars with
36\%, 30-36\%, and 23-25\% derived, respectively \citep[][]{Reynolds-1969}, using a comparable model that assumes no iron oxides in the mantle.}
More detailed studies strongly suggest that relative abundances of Fe, Mg, and Si are similar among the Sun, Earth, Mars, Venus, as well as meteorites \citep[e.g.,][]{Lodders-2003,Drake-2002,Lodders-1998,Khan-2008,Sanloup-1999}. Meteorites are believed to be chemically similar to the building blocks of planets \citep[][]{Morgan-1980}, because both are condensates from the solar nebula that experienced the same fractionation processes. In the solar system the Sun's photospheric relative Fe-abundance can thus be used as a proxy for the iron mass fraction for three out of four terrestrial planets. This fraction is in turn a key quantity for the characterization of exoplanets: from measurement of mass and radius alone, the interior composition can only be weakly constrained owing to model degeneracies \citep[][]{Dorn-2015}.

It is thus interesting to see whether for the planets analyzed by \citet[][]{Dressing-2015}, a compositional correlation between the host stars and their planets is possible, suggesting that it may not be restricted to the solar system case. This is predicted by planet formation models that include equilibrium condensation models corresponding to small distances to the host star ($\sim$1 AU). As a consequence, these relative abundances do not vary significantly within a disk, whether spatially and temporally \citep[e.g.,][]{Johnson-2012,Thiabaud-2014}\footnote{For a specific planet it is possible that an external process may have significantly altered its bulk composition like was the case for Mercury.}.  If we find observational evidence that many exoplanets follow a mass-radius relation that agrees with the relative composition in refractories of their host stars, this would mean that we can use stellar composition as a proxy to reduce the degeneracy of possible interior models, more specifically, by coupling structure and composition of core and mantle through bulk abundance ratios \citep[e.g.][]{Dorn-2015}.

With the goal of testing these possibilities, we present a detailed study of the chemical 
abundances for three out of five of the planets discussed in \citet[][]{Dressing-2015}: 
Kepler-10b, Kepler-93b, and CoRoT-7b. In Sects\,\ref{sec:observations} and \ref{sec:parameters} we present our
data, the derivation of stellar parameters, and chemical abundances. We then discuss the results in Sect.\,\ref{sec:conclusions}.

  \section{Observations and data}                               \label{sec:observations}

Spectra for \object{Kepler-10} and \object{Kepler-93} were obtained from the SOPHIE archive (OHP 1.93-m telescope).
The data was gathered as part of a program to derive masses and confirm the planetary nature of
detected Kepler candidates \citep[e.g.,][]{Santerne-2012}. The spectra cover the range between 3870 and 6940\AA. 
{The spectra were obtained using the high resolution mode (HR; R=75\,000) for Kepler-10, and the high efficiency mode (HE; R=40\,000) for Kepler-93.} In both cases Fiber\,B of the spectrograph was
pointed towards a sky region. The spectra were reduced using the SOPHIE pipeline. The sky spectra were 
subtracted from the stellar spectrum using the information available in Fiber\,B after correcting for the 
efficiency difference between the two fibers.
For Kepler-10 we combined five spectra obtained between 2011 July 7 and 2011 September 9 (other 
available spectra had very poor S/N). For Kepler-93 we combined two spectra obtained on the nights 
of 2014 March13 and 2014 April 26. The S/N of the final spectra is in the range of 80 and 60
as measured using continuum regions near 6500\AA, respectively.

For Kepler-10 we also gathered spectra obtained using the high resolution (R=115\,000) HARPS-N 
spectrograph \citep[][]{Cosentino-2012} -- courtesy of the HARPS-N consortium. We combined a total
of 110 high S/N spectra (S/N>50 in order 50) obtained between 2012 May 25 and 2013 October 18. 
The final spectrum has a S/N on the order of 600. A comparison between the results
obtained using the SOPHIE and HARPS-N spectra gives us an idea about the
possible systematic errors between parameters and abundances derived using different
spectra.

As we see below, we have already analyzed an existing HARPS spectrum for CoRoT-7.
We point the reader to \citet[][]{Mortier-2013b} for further details.

  \section{Stellar parameters and abundances}                           \label{sec:parameters}

Stellar parameters for Kepler-10 (independently {derived from} the SOPHIE and HARPS-N spectra) and Kepler-93 were derived 
in LTE, using a grid of plane-parallel, ATLAS9
model atmospheres \citep[][]{Kurucz-1993} and
the radiative transfer code MOOG
\citep[][]{Sneden-1973}. The methodology used is
described in detail in \citet[][]{Sousa-2011}. The full spectroscopic
analysis is based on the equivalent widths (EWs) of $\sim$250 Fe~{\sc i}
and 40 Fe~{\sc ii} weak lines by imposing ionization
and excitation equilibrium.
Parameters for CoRoT-7 have already been derived by our team (using the same tools
and methods) from 
a high resolution, high S/N HARPS spectrum \citep[][]{Mortier-2013b}. 

For FGK dwarfs the effective temperatures obtained
using this methodology were shown to be compatible
with estimates from recent applications of the
infrared flux method \citep[see, e.g.,][]{Santos-2013}.
Possible errors in the derived (always more uncertain) $\log{g}$ have little or no influence on the abundances of individual
species as derived from atomic (non-ionized) lines.

{The derived parameters for the three stars are listed in Table\,\ref{tab:param},
where we also list literature values for comparison.
It is interesting that the parameters and abundances derived for
Kepler-10 using the two spectra are similar despite the lower
S/N and resolution of the SOPHIE data. }

\begin{figure}
\begin{center}
\begin{tabular}{c}
\includegraphics[angle=270,width=1\linewidth]{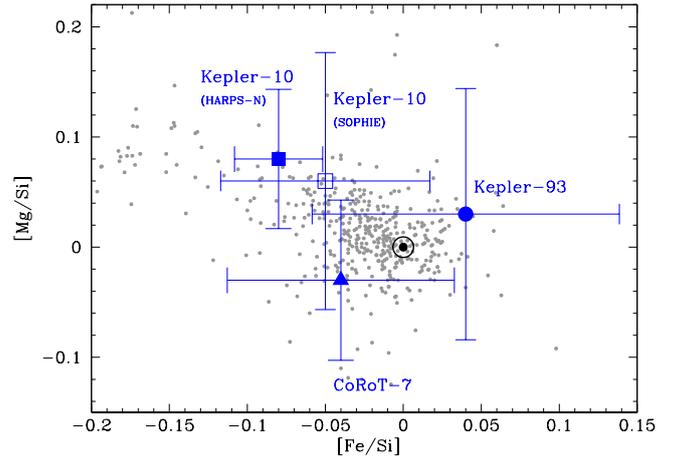}
\end{tabular}
\end{center}
\vspace{-0.5cm}
\caption{[Mg/Si] vs. [Fe/Si] for the studied stars, together with the comparison dwarf (gray small dots) from \citet{Adibekyan-2012b} 
with \emph{$T{}_{\mathrm{eff}}$} = \emph{T$_{\odot}$$\pm$$300$ K}. The position of the Sun is shown for comparison.}
\label{si_mg_fe}
\end{figure}

\subsection{Chemical abundances: Si, Mg, C, and O}

The abundances of magnesium (Mg) and silicon (Si) were derived for the three stars following the same methodology as in our previous works \citep[e.g.,][]{Adibekyan-2012,Adibekyan-2015}.
We used a LTE analysis relative to the Sun with the 2014 version of the code MOOG \citep[][]{Sneden-1973} and a grid of Kurucz ATLAS9 plane-parallel model 
atmospheres with no $\alpha$-enhancement.
For the EW measurements, we used ARES2 \citep{Sousa-2015}, for which the input parameters were the same as in \citet{Sousa-2011}.
A visual inspection was done to confirm the reliability of the ARES2 measurements.

The abundances for each star and element were calculated from the average of the values provided by all measured lines 
(3 lines for Mg and 16 lines for Si). A two-sigma clipping cut was performed to remove outliers.
Errors in the abundances of the elements [X/H] were evaluated adding quadratically the line-to-line scatter errors and errors 
induced by uncertainties in the model atmosphere parameters \citep[see, e.g.,][]{Adibekyan-2012}. 
The final abundances of Mg and Si are presented in Table\,\ref{tab:param}.
For CoRoT-7, abundances for Mg and Si perfectly match the ones derived in \citet[][]{Mortier-2013b} using an older version of
MOOG.

Carbon (C) and oxygen (O) abundances were derived with the same tools as Mg and Si 
but measuring the EWs of the lines with the task \textit{splot} in IRAF.
Atomic data for the O lines and for the Ni line-blend affecting the oxygen forbidden line 
at 6300$\AA$ were taken from \citet[][]{BertrandeLis-2015}. The data for the atomic carbon lines was taken from 
the VALD3 database\footnote{\url{http://vald.astro.univie.ac.at/~vald3/php/vald.php}}.

The final abundances derived using each individual line are shown in Table\,\ref{tab:CO}. The errors for the 
abundances are mainly due to the uncertainties in the position of the continuum. {In the error estimate, we also considered the
sensitivity of the C and O abundances due to the error in the atmospheric parameters, adding all terms quadratically as done for Mg and Si.}
The error caused by the continuum placement was estimated by recalculating the abundance using different continuum 
positions as measured by hand.
{Unfortunately, the S/N of the SOPHIE spectra is not high enough to allow us to measure the weak oxygen lines. Furthermore, the HARPS-N spectra for Kepler-10 shows telluric contamination in the region of the oxygen forbidden line. For CoRoT-7 the oxygen line at 6158\AA\ is too weak. Therefore we could only derive C/O ratios for Kepler-10 and CoRoT-7, based on a single oxygen indicator each.}

For reference, the solar values are also shown Table\,\ref{tab:CO}. While the carbon abundance in the Sun is well defined 
using the \ion{C}{I} atomic lines, the oxygen abundance is more uncertain, and the {different indicators (lines) provide different values} \citep[see discussion in][]{BertrandeLis-2015}. Therefore, in Table\,\ref{tab:CO} we give the oxygen abundance values individually. 
The [C/O] ratio is also computed using the corresponding solar abundance, depending on which oxygen line we could use.

\section{Discussion and conclusions}
\label{sec:conclusions}

In Fig.~\ref{si_mg_fe} we show the position of the stars in the [Mg/Si] - [Fe/Si] plane. The plot shows that
while Kepler-93 and CoRoT-7 do not seem to have significantly different abundance
ratios than the Sun, the situation for Kepler-10 seems different. Both the abundances derived from the SOPHIE
and HARPS-N spectra suggest that the star has a [Fe/Si] below solar ($-$0.08/$-$0.05 from HARPS-N and SOPHIE data, respectively).
The [Mg/Si] ratio observed in Kepler\,10 also seems to be above solar (0.08/0.06 from HARPS-N and SOPHIE data, respectively).
In this sense it is interesting to see that both sets of spectra provide very similar results, even if the whole analysis was
done in a fully independent way.
It is worth noting that since the model changes (variation in stellar parameters) induce similar effects in the abundances of these refractory elements 
(they partially cancel out when computing the ratios [X1/X2]), our reported errors are likely conservative. 

For Kepler-10 we find evidence for abundance ratios that are significantly different from solar,
though the differences are not higher than $\sim$0.1\,dex.
{For CoRoT-7 and Kepler-93, on the other hand, the abundance ratios seem to be very similar to solar.}
Following the galactic chemical evolution trends \citep[][]{Bensby-2005,Adibekyan-2012b}, we could expect that stars with
[Fe/H] below solar should on average present [Si/Fe] abundances that are slightly above solar 
(i.e., [Fe/Si] ratios below solar), though perfectly allow the opposite trend given the observed dispersions. 

Concerning C and O, we first note that the carbon abundances for Kepler-10, as derived from different spectra (HARPS-N and SOPHIE), agree well within the errors. 
The [C/O] ratio for this star seems to be subsolar. This is expected since oxygen, as a pure $\alpha$-element, grows more rapidly 
than carbon as a function of [Fe/H], for [Fe/H] $<$ 0 \citep[e.g.,][]{BertrandeLis-2015,Bensby-2005}. In any case, the
observed [C/O] ratio values are not expected to have a significant impact on the bulk composition of the planet \citep[][]{Bond-2010,Delgado-Mena-2010}
when compared to a solar ratio.

On the other hand, for CoRoT-7 we could expect a [C/O] ratio that is more similar to the solar value
owing to its more similar metallicity. However, its elevated oxygen abundance yields a lower [C/O] ratio. Although the determination of oxygen abundances can often be problematic for cooler 
stars such as CoRoT-7, in this case the quality of the spectrum 
allowed us to measure the EW of the forbidden line perfectly, and the spectrum does not seem to show telluric contamination 
in the region (which could produce an artificial enhanced oxygen abundance). We note that the high error of [C/O] for 
this star precludes firm conclusions. 

\begin{table}[t!]
\caption{Mass fractions of heavy element, total fraction of heavy elements, and iron mass fraction among refractory species (values in \%).}
\label{tab:massfract}
\begin{tabular}{lccccc}
\hline
\noalign{\medskip} 
Quantity &  C-7 & K-93 & K-10 & Sun & Sun$^f$\\\hline
\noalign{\medskip}
H$_{2}$O\tablefootmark{a} & 0.75$\pm$0.31 & 0.54$\pm$0.22 &0.98$\pm$0.20 & 0.50 & 0.51  \\   
CH$_{4}$\tablefootmark{a} & 0.32$\pm$0.05 & 0.35$\pm$0.13 &0.36$\pm$0.04 & 0.37 & 0.29  \\
Fe\tablefootmark{a}              & 0.14$\pm$0.01 & 0.09$\pm$0.01 &0.10$\pm$0.00 & 0.13 & 0.17\tablefootmark{d}  \\
MgSiO$_{3}$\tablefootmark{a} & 0.25$\pm$0.08 & 0.10$\pm$0.06 &0.11$\pm$0.05 & 0.19 & \multirow{2}{*}{0.27\tablefootmark{e}}  \\    
Mg$_{2}$SiO$_{4}$\tablefootmark{a}  & 0.05$\pm$0.06 & 0.08$\pm$0.06 &0.14$\pm$0.06 & 0.08 &   \\\hline
Z\tablefootmark{b}& 1.50$\pm$0.31 & 1.17$\pm$0.25 &1.69$\pm$0.21 & 1.26 & 1.32  \\ \hline
$f_{\rm iron}$\tablefootmark{c} & 31.6$\pm$2.6 & 34.7$\pm$3.7 & 27.5$\pm$1.7 & 33.2 & 38.0  \\
\hline
\end{tabular}
\newline
\tablefoottext{a}{The $m_{\rm H2}$ and $m_{\rm He}$ are 
between 74.7-75.1\% and 23.6-23.7\%, respectively.}
\tablefoottext{b}{Summed mass percent of all heavy elements.}
\tablefoottext{c}{$m_{\rm Fe}/(m_{\rm Fe}+m_{\rm MgSiO3}+m_{\rm Mg2SiO4})$.}
\tablefoottext{d}{Includes all metal species and FeS.}
\tablefoottext{e}{Includes all silicates and oxides.}
\tablefoottext{f}{\citet[][]{Lodders-2003}.}
\end{table}

Whether the observed compositional differences are relevant when modeling the interior structure of the exoplanets is the next obvious question to answer. 
In particular, we want to know whether the interiors structure of the planets presented by \citet[][]{Dressing-2015} can be explained when using the specific stellar abundances as proxies for the planet bulk compositions. Given the large uncertainties on mass and radius, as well as on the abundances, we suppose that (1) this is likely the case and (2) the single model of an Earth-like iron-core mass fraction as used by \citet[][]{Dressing-2015} lies within the generally wide range of possible interior models.
A detailed  analysis of this including a full error analysis will be presented in future theoretical work. Here we only make first simple considerations regarding the relative expected abundances of the most important rock-forming elements and, in particular, of iron.

We only have abundances of Fe, Si, Mg, O, and C. However, these rocky-forming elements, together with H and He, are the most relevant ones for controlling the species expected from equilibrium condensation models, \citep{Lodders-2003,Seager-2007}, namely H$_{2}$, He, CH$_{4}$, H$_{2}$O, Fe, MgSiO$_{3}$, and Mg$_{2}$SiO$_{4}$ (All stars have N$_{\rm Mg}>N_{\rm Si}$). A simplified model for the expected mass fractions of different compounds using these species is thus a reasonable approach. In this chemical network, the molecular abundances and therefore the mass fraction can be found from the atomic abundances with simple stoichometry:
\vspace{-0.15truecm}
\begin{eqnarray}
N_{\rm O}&=&N_{\rm H_2O}+3 N_{\rm MgSiO_3}+4 N_{\rm Mg_2SiO_4}\\[-0.1truecm]
N_{\rm Mg}&=&N_{\rm MgSiO_3}+2  N_{\rm Mg_2SiO_4}\\[-0.1truecm]
N_{\rm Si}&=&N_{\rm MgSiO_3}+  N_{\rm Mg_2SiO_4}
.\end{eqnarray}
Similar equations hold for the other species. This way we calculated the expected mass fractions, in particular $f_{\rm iron}$, for the three stars analyzed in this paper. Table\,\ref{tab:massfract} shows the result in terms of mass fractions $m_{\rm X}$ of the five heavy element species  (in percentage points), the total mass fraction of heavy elements $Z$, and finally the iron mass fraction among the refractory elements, which is the key result. 
In the table we also showed, for comparison, the results with our basic chemistry model for the Sun using the solar composition of \citet{Asplund-2009}. The table also contains the condensate abundances derived  by \citet{Lodders-2003} (her Table 11, photospheric values) with a full condensation model that includes a much higher number of species. Kepler-93 (K-93) has no oxygen measurement. We assumed a solar oxygen abundance for this star. This is a good assumption given the expected oxygen-to-iron ratio for stars with such metallicity \citep[e.g.,][]{BertrandeLis-2015}. {Error bars were derived as explained in Appendix\,\ref{sec:errors}.}

The table shows that owing to the approximately scaled solar composition of CoRoT-7 (C-7) and Kepler-93 (under the assumption of solar $N_{\rm O}$), their $f_{\rm iron}$ is very similar to the one of the Sun (32-35\%). Only for Kepler-10 (K-10), which is more depleted in Fe than in Mg and Si relative to the Sun, do we see a slightly lower $f_{\rm iron}=27.5$\% (and higher H$_2$O), as expected. Curiously, both Kepler-10b and Kepler-10c seem to be ``oversized'' in the mass-radius diagram of \citet[][]{Dressing-2015} when compared to the Earth composition model. In any case, we find that for the three planets analyzed by \citet[][]{Dressing-2015}, the iron mass fraction inferred from the mass-radius relationship ($\sim$17\%) is on a $\sim$10\%-20\% level, in agreement with the iron abundance derived here from the host star photospheric composition, {and in even better agreement with the mass fraction ($\sim$29\%-32\%) expected from more detailed models of solar system planets \citep[for the Earth see, e.g.,][]{McDounough-1995}}. If confirmed, this interesting result doubles the number of rocky planets (including the solar system's) where the mass of the iron core can be estimated if the stellar composition is known, and brings it to six out of seven rocky planets where such an inference can be made (the exception is Mercury). As discussed in the introduction, this is important for reducing the compositional degeneracy of planet interiors \citep{Dorn-2015} and for constraining impact events.

A much more detailed analysis that combines host star abundances, mass and radius measurements, and internal structure calculations should be done in future work, taking all the associated measurement and model uncertainties into account. If the agreement of the planetary core mass (derived from the $M-R$ relationship) and the stellar photospheric abundances found here is confirmed and extended to additional planets, this would allow a better characterization of exoplanet interiors, even if a significantly different composition (as for Mercury) cannot be excluded a priori for an individual planet. This shows that a detailed chemical analysis of exoplanet hosts is important not only for the architecture of a planetary system, but also for the interior structure, composition, and potential habitability of individual planets. 
The detailed characterization of planets discovered and characterized by future space missions such as TESS, CHEOPS, and PLATO-2.0 will certainly gain from this analysis.

\begin{acknowledgements}
We would like to thank the anonymous referee for the useful comments and suggestions. This work was supported by Funda\c{c}\~ao para a Ci\^encia e a Tecnologia (FCT) through the research grant UID/FIS/04434/2013. PF, NCS, and SGS also acknowledge support from FCT through Investigador FCT contracts of reference IF/01037/2013, IF/00169/2012, and IF/00028/2014, and POPH/FSE (EC) by FEDER funding through the program ``Programa Operacional de Factores de Competitividade - COMPETE''. PF acknowledges support from FCT in the form of project reference IF/01037/2013CP1191/CT0001. AS is supported by the EU under a Marie Curie Intra-European Fellowship for Career Development with reference FP7-PEOPLE-2013-IEF, number 627202. EDM and VA acknowledge the support from FCT in form of the grants SFRH/BPD/76606/2011 and SFRH/BPD/70574/2010. This work results within the collaboration of the COST Action TD 1308. CM acknowledges the support from the Swiss National Science Foundation under grant BSSGI0\_155816. AM received funding from the European Union Seventh Framework Program (FP7/2007-2013) under grant agreement number 313014 (ETAEARTH).

\end{acknowledgements}

\bibliographystyle{aa}
\bibliography{santos_bibliography}


\Online
\begin{appendix}

\section{Stellar parameters and chemical abundances}

\begin{table*}[h!]
\caption{Stellar atmospheric parameters and iron, magnesium, and silicon abundances for our targets.}
\label{tab:param}
\centering
\begin{tabular}{lccccccll}
\hline\hline
Star & T$_{eff}$ & $\log g$ & $\xi$ & [Fe/H] & [Mg/H] & [Si/H] & Source &  Instrument\\
        & (K)            & (dex)      & (km s$^{-1}$) & (dex) & (dex) &(dex) & & \\
\hline
Kepler-10     &  5671$\pm$17   &   4.34$\pm$0.01 &   0.92$\pm$0.02  &  $-$0.14$\pm$0.02 & +0.02$\pm$0.06 & $-$0.06$\pm$0.02 & This paper & HARPS-N \\
Kepler-10     &  5614$\pm$39  &    4.27$\pm$0.04  &  0.82$\pm$0.05  &  $-$0.14$\pm$0.03 & $-$0.03$\pm$0.10 & $-$0.09$\pm$0.06 & This paper & SOPHIE \\
Kepler-93     &  5689$\pm$48   &   4.56$\pm$0.06  &  1.02$\pm$0.07  &  $-$0.15$\pm$0.04 & $-$0.16$\pm$0.07 & $-$0.19$\pm$0.09 & This paper & SOPHIE \\
CoRoT-7     &  5288$\pm$27   &   4.40$\pm$0.07  &  0.90$\pm$0.05  &  $+$0.02$\pm$0.02 & $+$0.03$\pm$0.03 & $+$0.07$\pm$0.08 & \citet[][]{Mortier-2013b} & HARPS \\
\hline
\multicolumn{9}{l}{Literature values:}\\
Kepler-10   &   5627$\pm$44 & 4.34$\pm$0.01 &                                        &  $-$0.15$\pm$0.04 &                                     &                                   &\citet[][]{Batalha-2011} &  \\
                     & 5721$\pm$26  & 4.34  &                                           & $-$0.14$\pm$0.02 &                                     &                                   &\citet[][]{Dumusque-2014} &  \\
Kepler-93   &   5669$\pm$75 & 4.47$\pm$0.01 &                                        &  $-$0.18$\pm$0.10 &                                     &                                   &\citet[][]{Ballard-2014} &  \\
CoRoT-7   &    5250$\pm$60 & 4.47$\pm$0.05 &                                        &      $+$0.12$\pm$0.06 &                                   &                                   &\citet[][]{Bruntt-2010} &  \\
                   &    5275$\pm$75 & 4.50$\pm$0.10 &                                        &      $+$0.03$\pm$0.06 &                                   &                                   &\citet[][]{Leger-2009} &  \\
\hline
\end{tabular}
\end{table*}

\begin{table*}
\caption{Carbon and oxygen abundances (in dex) for our stars and the Sun, using the Kurucz Solar Atlas (Kurucz et al. 1984).}
\label{tab:CO}
\begin{tabular}{lccccccccc}
\hline\hline
\noalign{\medskip} 
Star & logC$_{5052\AA}$ & logC$_{5380\AA}$ & logO$_{6158\AA}$& logO$_{6300\AA}$ & [C/H]  & [O/H] &  [C/O] & C/O$^e$ & Mg/Si$^e$ \\
\noalign{\medskip} 
\hline
\noalign{\medskip} 
Kepler-10\tablefootmark{b} & 8.48 & 8.47 & 8.92 & ---- & -0.01$\pm$0.05 & 0.21$\pm$0.08 & -0.22 $\pm$0.10& 0.36 & 1.48 \\   
Kepler-10\tablefootmark{a} & 8.52 & 8.44 & ---- & ---- &  0.00$\pm$0.08 & ----          & ----           & ---- & 1.41 \\
Kepler-93\tablefootmark{a} & 8.35 & 8.59 & ---- & ---- & -0.02$\pm$0.16 & ----          & ----           & ---- & 1.32 \\
CoRoT-7\tablefootmark{c}   & 8.38 & 8.48 & ---- & 8.83 & -0.06$\pm$0.07 & 0.18$\pm$0.15 & -0.24$\pm$0.17 & 0.40 & 1.12 \\    
Sun                        & 8.48 & 8.49 & 8.71 & 8.65 &  0.00          & 0.00          &  0.00      & 0.64\tablefootmark{d} & 1.23 \\
\hline
\end{tabular}
\newline
\tablefoottext{a}{SOPHIE}
\tablefoottext{b}{HARPS-N}
\tablefoottext{c}{HARPS}
\tablefoottext{d}{C/O ratios derived using the average values for the oxygen and carbon abundances given by different lines.}
\tablefoottext{e}{$A/B  = N_{A} / N_{B} = 10^{\log \varepsilon(A)}/10^{\log \varepsilon(B)}$, where $\log \varepsilon(A)$ and $\log \varepsilon(B)$ are the absolute abundances.}
\end{table*}

\section{Errors}
\label{sec:errors}

The errors in the mass fractions were calculated using a Monte Carlo approach. We
randomly drew 10$^5$ values of the derived stellar abundances abundances following
a Gaussian distribution centered on the derived abundance and with a sigma corresponding
to the derived errors as listed in Tables\,\ref{tab:param} and \ref{tab:CO}.
For each drawn set of [Si/H], [Mg/H], [O/H], and [C/H], we then computed the expected 
fractions. The distribution of the resulting values allowed us to derive the one-sigma error,
as listed in Table\,\ref{tab:massfract}.

{We note that this procedure was done in a self consistent way. All the mass fractions
were computed simultaneously for each set of input abundances. This is important because the
resulting mass fraction values are correlated between themselves.}

For Kepler-93, since the oxygen abundance was assumed to be solar (no O abundance could
be derived from our data), we considered that the [O/H] has an uncertainty of 0.15\,dex.

\end{appendix}

\end{document}